\documentclass[11pt]{article}

\input epsf
\usepackage{latexsym}
\usepackage{amssymb}
\usepackage{amsmath}
\usepackage[dvips]{graphicx}
\usepackage{array}

\begin{document}
\centerline{\Large \bf  Large scale magnetic fields from gravitationally coupled electrodynamics}

\vskip 2 cm

\centerline{Kerstin E. Kunze
\footnote{E-mail: kkunze@usal.es} }

\vskip 0.3cm

\centerline{{\sl Departamento de F\'\i sica Fundamental,}}
\centerline{{\sl Universidad de Salamanca,}}
\centerline{{\sl Plaza de la Merced s/n, E-37008 Salamanca, Spain }}

\vskip 1.5cm

\centerline{\bf Abstract}
\vskip 0.5cm
\noindent
The generation of primordial magnetic seed fields during inflation is studied in a theory derived from the one-loop vacuum polarization effective action of the photon in a curved background.
This includes terms which couple the curvature to the Maxwell tensor. 
The resulting magnetic field strength is estimated in a model where the inflationary phase is directly matched to the standard radiation dominated era.
The allowed parameter region is analyzed and compared with the bounds necessary to seed the galactic magnetic field. It is found that magnetic fields of cosmologically interesting field strengths can be generated.

\vskip 1cm

\section{Introduction}
\setcounter{equation}{0}

Magnetic fields are found to be associated with nearly all structures in the universe.
They are observed on stellar upto possible supercluster scales \cite{mag}.
In some processes such as star formation they can play an important role.
Equally the physics of cosmic rays indicate the existence of a large scale galactic magnetic field.
Furthermore, ultra high energy cosmic rays which are most likely of extra galactic origin could be used to study the properties of galactic and extra galactic magnetic fields \cite{cr}.

As to the origin of the observed magnetic fields there seem to be generally 
speaking two broad classes of mechanisms \cite{mag}. On the one hand there are battery-type mechanisms which work on the basis of charge separation which leads to a current and finally induces a magnetic field. On the other hand there are dynamo type mechanisms which amplify an initial seed magnetic field.
The former have a coherence scale of the order of the domain associated with the battery-mechanism, which in a cosmological context is always smaller than the horizon at the epoch of creation. 
The latter involves magnetic fields whose correlation length is of the order of the 
region of interest, which in the case of a galactic magnetic field, would be a proto galactic scale of the order  of  1 Mpc today which requires a mechanism to create magnetic fields with rather large coherence scales.
A natural mechanism for this is provided by the amplification of perturbations of the electromagnetic field during inflation. Since quantum perturbations are stretched beyond the horizon during inflation and upon leaving the causal domain becoming classical, there is no problem with the coherence length. However, in general, in a background geometry with flat spatial section the resulting field strength of the primordial magnetic field after inflation in standard electrodynamics is far too small to, for example, serve as a seed field for a potential galactic dynamo explaining the observed  galactic magnetic field of  the order of  $10^{-6}$ G. 
Starting with \cite{tw} this motivated an intensive study of alternative models of some kind of coupling of linear electrodynamics to either other fields in the theory, such as scalar fields 
\cite{mag-sc} or gravity \cite{tw,mgrav} including extra dimensions \cite{mag-ex}.
Other models break explicitly Lorentz invariance considering a non zero photon mass \cite{tw,BLI}.
Recently there has also been interest in magnetic field generation within models of nonlinear electrodynamics which naturally occur when quantum corrections and self couplings of the electromagnetic field are taken into account \cite{nled}.
Furthermore, there are models within the standard model or its
supersymmetric extensions \cite{other}.
 In the case of models with curved spatial sections 
it was shown that even in the minimally coupled model of linear electrodynamics 
strong enough magnetic fields can be created \cite{tk}.

Here we are returning to a model where the 
electromagnetic field is gravitationally coupled which first has been proposed in the context of the generation of primordial magnetic fields during inflation in \cite{tw}.
In particular the model under consideration derives form the one loop effective action 
of vacuum polarization in QED in a gravitational background \cite{dh}. 
As shown in \cite{dh}  this leads to birefringence of the electromagnetic wave where the photons have  velocities depending on their polarization which can exceed the speed of light. This last observation leads to an interesting causal structure of these space-times \cite{gs}.

\section{Estimating the magnetic field strength}
\setcounter{equation}{0}
In \cite{dh} it was shown that to lowest order  
the propagation of a photon in a gravitational background 
 is described by the Lagrangian \cite{dh,tw}
\begin{eqnarray}
{\cal L}=-\frac{1}{4}F_{\mu\nu}F^{\mu\nu}-\frac{1}{4m_e^2}\Big[
bR F_{\mu\nu}F^{\mu\nu}+cR_{\mu\nu}F^{\mu\kappa}F^{\nu}_{\;\;\kappa}
+dR_{\mu\nu\lambda\kappa}F^{\mu\nu}F^{\lambda\kappa}+f(\nabla_{\mu}F^{\mu\nu})(\nabla_{\alpha}F^{\alpha}_{\;\;\nu})
\Big]
\end{eqnarray}
The expansion parameter is bascially the square of the Compton wave length of the electron which enters due to the fact that vacuum polarization effectively gives the photon a non zero "size" due to electron positron pair creation. The parameters $b$, $c$, $d$, and $f$ are are free parameters, which, however, have been calculated in \cite{dh} in the weak gravitational field limit.
It was found that the coefficient of the last term is of the order $e^2$ and can therefore be neglected.
Moreover, unless $f$ is to be chosen much larger than the other coefficients it is expected to be negligible in general, as  can be appreciated from the equations of motion,
\begin{eqnarray}
\nabla^{\mu}F_{\mu\nu}&+&\frac{1}{m_e^2}\nabla^{\mu}\Big[bRF_{\mu\nu}+\frac{c}{2}\Big(
R^{\lambda}_{\;\;\mu}F_{\lambda\nu}-R^{\lambda}_{\;\;\nu}F_{\lambda\mu}\Big)+dR^{\lambda\kappa}_{\;\;\;\;\;\;\mu\nu}F_{\lambda\kappa}\Big]\nonumber\\
&+&\frac{f}{2m_e^2}
\left(
\nabla_{\alpha}\nabla^{\alpha}\nabla^{\beta}F_{\beta\nu}+
R^{\alpha}_{\;\;\nu}\nabla^{\beta}F_{\beta\alpha}
\right)=0.
\label{e1}
\end{eqnarray}
Thus the new type of term that the last term contributes involves higher derivatives which are suppressed by a factor $\sim 1/\sigma^2$, where $\sigma$ is a typical scale. Therefore, in the following this term will neglected and $f\equiv 0$.

The Maxwell tensor is expressed in terms of the gauge potential $A_{\mu}$, that is $F_{\mu\nu}=\partial_{\mu}A_{\nu}-\partial_{\nu}A_{\mu}$ and the Coulomb gauge is used
$A_0=0$,  $\partial_iA_i=0$.
Furthermore, the background cosmology is described by the line element,
\begin{eqnarray}
ds^2=a^2(\eta)\Big(-d\eta^2+dx^2+dy^2+dz^2\Big),
\label{e2}
\end{eqnarray}
where $a(\eta)$ is the scale factor describing  a model, in which the inflationary stage is directly matched at $\eta=\eta_1$ to the standard radiation dominated era,
\begin{eqnarray}
a(\eta)=\left\{
\begin{array}{lr}
a_1\left(\frac{\eta}{\eta_1}\right)^{\beta}& \eta<\eta_1\\
&\\
a_1\left(\frac{\eta-2\eta_1}{-\eta_1}\right)&\eta\geq\eta_1.
\end{array}
\right.
\end{eqnarray}
In the following $a_1\equiv 1$.
For $\beta=-1$ de Sitter inflation is realized and for $-\infty<\beta<-1$ the model has a stage of power law inflation for $\eta<\eta_1$. The exponent $\beta$ is related to the equation of state of matter, defined by,
$p=\gamma\rho$, by $\beta=\frac{2}{3\gamma+1}$. Note that $\gamma=\frac{2-\beta}{3\beta}$ takes values between -1 and $-\frac{1}{3}$.
The Fourier expansion of the gauge potential is given by, 
\begin{eqnarray}
A_j(\eta,\vec{x})=\int \frac{d^3k}{(2\pi)^{\frac{3}{2}}\sqrt{2k}}\sum_{\lambda=1}^{2}\epsilon_{\vec{k}\;j}^{(\lambda)}\left[a^{(\lambda)}_{\vec{k}}A_k(\eta)e^{i\vec{k}\cdot\vec{x}}+a^{(\lambda)\,\dagger}_{\vec{k}}A^*_k(\eta)e^{-i\vec{k}\cdot\vec{x}}\right],
\end{eqnarray}
where the sum is over the two polarization
 states and  $\epsilon_{\vec{k}\;j}^{(\lambda)}$ are the polarization vectors satisfying,
 $\vec{\epsilon}_{\vec{k}}^{\;(\lambda)}\cdot\vec{k}=0$. Furthermore, the amplitude $A_k$ satisfies the same mode equation for both polarization states. Thus the index  $(\lambda)$ is suppressed
 in $A_k$.

In Fourier space in the background model (\ref{e2}) equation (\ref{e1}) results in 
\begin{eqnarray}
F_1(\eta)\ddot{A}_k+F_2(\eta)\dot{A}_k+F_3(\eta)k^2A_k=0,
\end{eqnarray}
where a dot indicates $\frac{d}{d\eta}$ and   
\begin{eqnarray}
F_1(\eta)&=&1+\frac{\mu_1}{m_e^2\eta_1^2}\left(\frac{\eta}{\eta_1}\right)^{-2(\beta+1)}
\hspace{2cm}
\mu_1=\beta\Big[6b(\beta-1)+c(\beta-2)-2d\Big]
\nonumber\\
F_2(\eta)&=&\frac{\mu_2}{\eta_1^3m_e^2}\left(\frac{\eta}{\eta_1}\right)^{-2\beta-3}
\hspace{2.85cm}
\mu_2=-2(\beta+1)\mu_1
\nonumber\\
F_3(\eta)&=&1+\frac{\mu_3}{\eta_1^2m_e^2}\left(\frac{\eta}{\eta_1}\right)^{-2(\beta+1)}
\hspace{2cm}
\mu_3=\beta\Big[6b(\beta-1)+c(2\beta-1)+2d\beta\Big].
\end{eqnarray}
The standard quantization procedure requires the corresponding action of the field to be diagonal. This can be achieved by using the canonical field $\Psi$ defined by
\begin{eqnarray}
\Psi=F_1^{\frac{1}{2}}A_k
\end{eqnarray}
which satisfies the mode equation,
\begin{eqnarray}
\Psi''+P\Psi=0,
\label{e3}
\end{eqnarray}
where a new dimensionless variable $z\equiv -k\eta$ has been defined and $' \equiv \frac{d}{dz}$.
Moreover,
\begin{eqnarray}
P=\frac{1}{4}\frac{\kappa_1z^{-4\beta-6}}{\Big[1+\kappa_2z^{-2(\beta+1)}\Big]^2}+\frac{1}{2}
\frac{\kappa_3z^{-2\beta-4}}{1+\kappa_2z^{-2(\beta+1)}}
+\frac{1+\kappa_4z^{-2(\beta+1)}}{1+\kappa_2z^{-2(\beta+1)}},
\end{eqnarray}
and 
\begin{eqnarray}
\kappa_1&\equiv&\mu_2^2\kappa_0^2\hspace{1.5cm}
\kappa_2\equiv\mu_1\kappa_0\hspace{1.5cm}
\kappa_3\equiv(2\beta+3)\mu_2\kappa_0\hspace{1.5cm}
\kappa_4\equiv\mu_3\kappa_0\nonumber\\
{\rm where}\hspace{0.5cm}
\kappa_0&\equiv&\left(\frac{m_e}{H_1}\right)^{-2}\left(\frac{k}{k_1}\right)^{2(\beta+1)}.
\end{eqnarray}
In deriving these expressions the maximally amplified (comoving) wavenumber $k_1$ has been 
defined by $k_1\equiv\frac{1}{|\eta_1|}$. $H_1$ is the value of the Hubble paramter at the beginning of the radiation dominated stage at $\eta_1$. It is related to $k_1$ by $k_1\sim H_1$.
The spectrum of the resulting magnetic field is determined by calculating the Bogoliubov coefficients which connect the "in" and "out" vacua. 
The relevant solutions of the mode equation (\ref{e3}) are those on superhorizon scales, $z\ll 1$, in which case equation (\ref{e3}) reduces to
\begin{eqnarray}
\Psi''+(\xi_1z^{-2}+\xi_2)\Psi=0,
\label{e4}
\end{eqnarray} 
where 
\begin{eqnarray}
\xi_1=-(\beta+1)(\beta+2)
\hspace{2cm}
\xi_2=\frac{6b(\beta-1)+c(2\beta-1)+2d\beta}
{6b(\beta-1)+c(\beta-2)-2d}.
\end{eqnarray}
For $\beta=-1$, which describes de Sitter inflation, $\xi_1=0$ and $\xi_2=1$  the solution is a plane wave which was also noted in \cite{dh,tw}. Furthermore,  $\beta=-2$  also gives $\xi_1=0$, but $\xi_2=\frac{18b+5c+4d}{18b+4c+2d}$.
 Solving equation (\ref{e4}) during power law inflation, $\beta<-1$ and $\beta\neq -2$, results in the following solution in terms of the Hankel function of the second kind, $H_{\nu}^{(2)}(x)$, 
 \begin{eqnarray}
 \Psi^{\rm (I)}=\sqrt{\frac{\pi}{2k}}\sqrt{z}H_{\nu}^{(2)}(\sqrt{\xi_2}\,z),
\hspace{1cm}{\rm where}
\hspace{0.5cm}
\nu=\left|\beta+\frac{3}{2}\right|
 \end{eqnarray}
which gives the correctly normalized incoming wave function for $\eta\rightarrow-\infty$ for $\xi_2>0$ .
During the radiation dominated stage the additional curvature terms can be neglected and thus the mode equation (\ref{e4}) simplifies to that of a free harmonic oscillator which is solved by the superposition of plane waves,
\begin{eqnarray}
\Psi^{\rm (RD)}=\frac{1}{\sqrt{k}}\left(c_{+}e^{-i(z-z_1)}+c_{-}e^{i(z-z_1)}\right),
\end{eqnarray} 
where $z_1\equiv k|\eta_1|$ and $c_{\pm}$ are the Bogoliubov coefficients.
Since the aim here is to calculate the magnetic field energy spectrum at galactic scale which re-enters during the radiation era it is enough to only consider the radiation dominated stage. Matching the solutions of the  gauge potential and its first derivatives at $\eta=\eta_1$ on superhorizon scales determines $c_+$ and $c_{-}$. In particular, using the small argument limit of the Hankel functions \cite{as} 
$|c_{-}|^2$ is found to be, for $\beta\neq-\frac{3}{2}$ 
\begin{eqnarray}
|c_{-}|^2\simeq\frac{\left[\Gamma(\nu)\right]^2}{8\pi\mu_1}\left(\frac{1}{2}-\nu\right)^2
\left(\frac{m_e}{H_1}\right)^2\left(\frac{\xi_2}{4}\right)^{-\nu}\left(\frac{k}{k_1}\right)^{-1-2\nu}
\end{eqnarray}
where it was used that in the approximation used here,
$F_1(\eta_1)\simeq\mu_1\left(\frac{m_e}{H_1}\right)^{-2}$. In the case $\beta=-\frac{3}{2}$
the limiting behavior of the mode function on superhorizon scales leads to a 
a divergent  factor $\ln^2\left(\sqrt{\xi_2}\frac{k}{k_1}\right)$ in $|c_{-}|^2$. Therfore, this case will not be considered further.
Including both polarization states the total spectral energy density of the photons is given by (cf., e.g., \cite{dks})
\begin{eqnarray}
\rho(\omega)\equiv\frac{d\rho}{d\log k}\simeq
2\left(\frac{k}{a}\right)^4\frac{|c_{-}|^2}{\pi^2}
\label{e5}
\end{eqnarray}
Since the electric field decays rapidly due to the high conductivity of the radiation dominated universe,  the spectral energy density (\ref{e5}) gives a measure of the magnetic field energy density, $\rho_{\rm B}$, which results in the commonly used ratio of magnetic field over background radiation energy density $r\equiv\frac{\rho_{\rm B}}{\rho_{\gamma}}=\frac{\Omega_{\rm B}}{\Omega_{\gamma}}$ \cite{tw}, where the density parameter of radiation is given by
$\Omega_{\gamma}= \left(\frac{H_1}{H}\right)^2\left(\frac{a_1}{a}\right)^4$ , for $\beta\neq -2,-\frac{3}{2}, -1$,
\begin{eqnarray}
r\simeq
\frac{2\left[\Gamma(\nu)\right]^2}{3\pi^2\mu_1}\left(\frac{1}{2}-\nu\right)^2\left(
\frac{m_e}{M_{\rm P}}\right)^2\left(\frac{\xi_2}{4}\right)^{-\nu}\left(\frac{k}{k_1}\right)^{3-2\nu}
\label{e6}
\end{eqnarray}
where $M_{\rm P}$ is the Planck mass.
An  expression similar to (\ref{e6}) is obtained when the magnetic field energy density is calculated using the two point function of the magnetic field, $\langle B_i(\vec{k})B_j^*(\vec{k}')\rangle$.
The form of the magnetic field spectrum (\ref{e6}) imposes the constraint $\nu\leq\frac{3}{2}$  which implies the range for $\beta$ given by  $-3<\beta<-1$ taking into account the constraint from power law inflation.

\begin{figure}[ht]
\centerline{\epsfxsize=2.2in\epsfbox{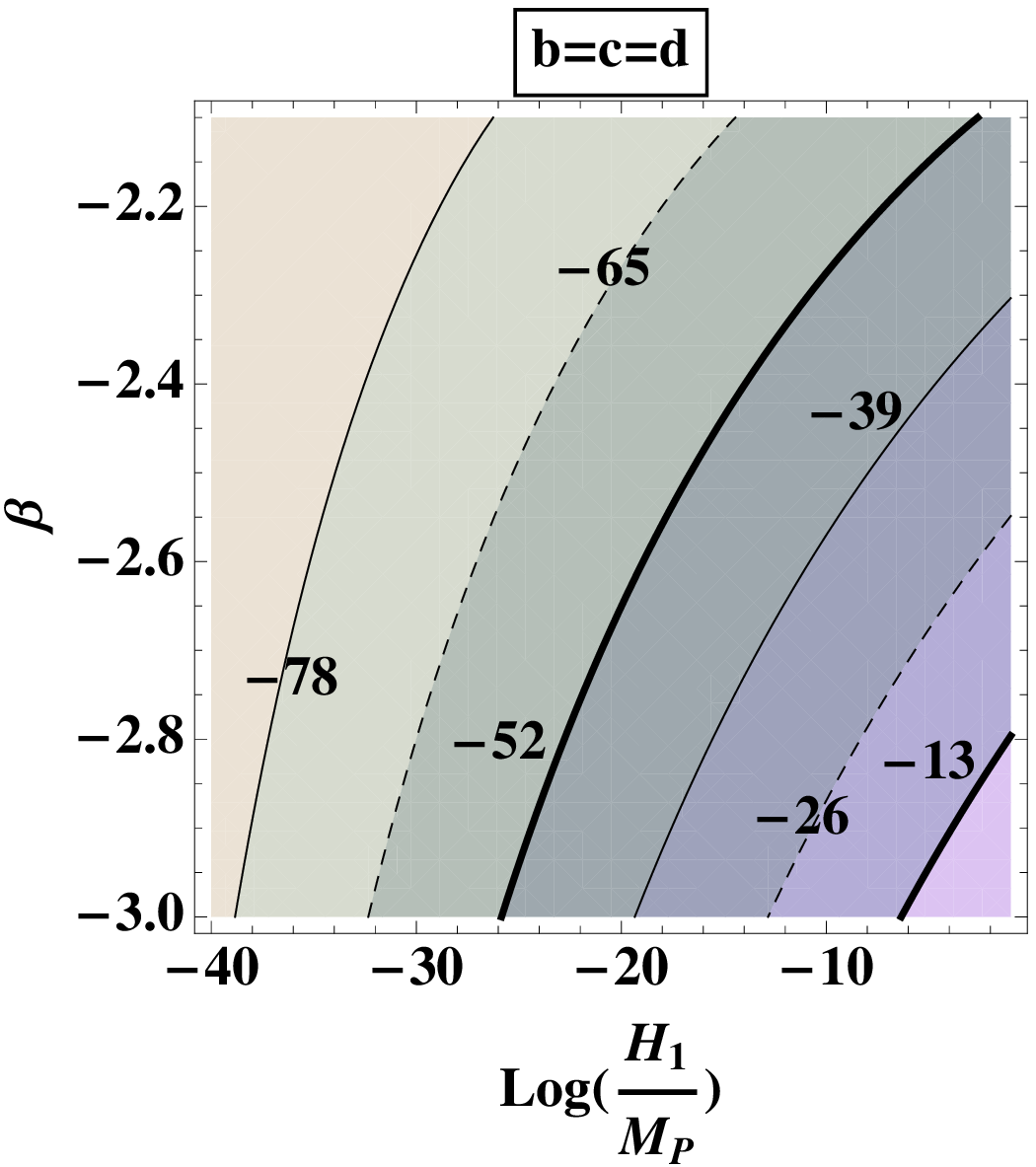}\hspace{2cm}
\epsfxsize=2.2in\epsfbox{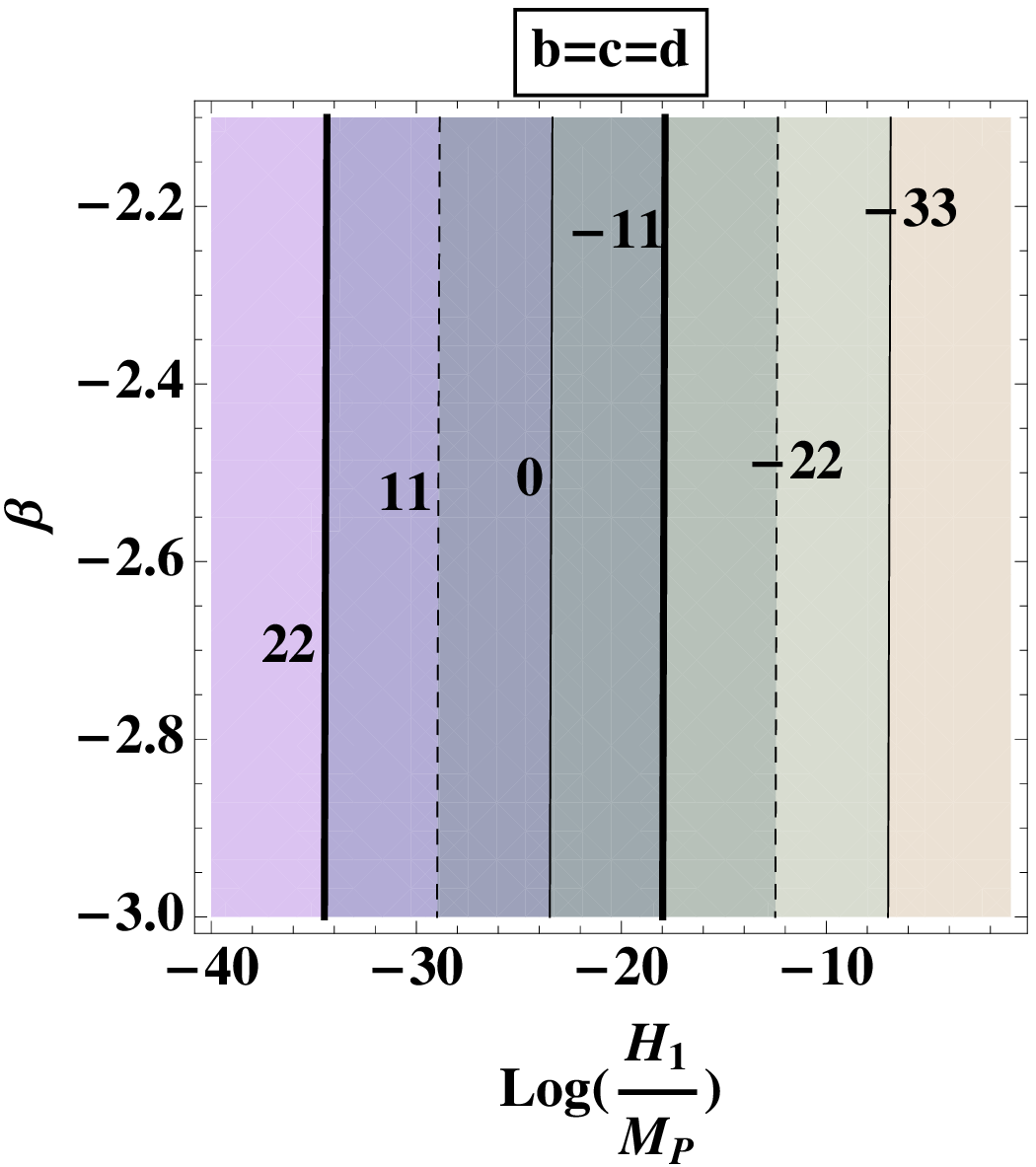}
}
\caption{{\it Left:} 
The contour lines for the maximum value of the logarithm of the ratio of magnetic to background radiation energy density is shown, that is 
$\log_{10} r_{\rm max}$, for  $b=c=d$.
The values of $\log\left(\frac{H_1}{M_{\rm P}}\right)$  correspond to
reheat temperatures between 0.1 GeV and $10^{19}$ GeV.
 The numbers within the graph refer to the value of $\log_{10}r_{\rm max}$ along the closest contour line.
\newline
{\it Right:}
The contour lines for the minimum value of the logarithm of the parameter $b$, that is 
$\log_{10}b_{\rm min}$, are shown in the model $b=c=d$.
 The numbers within the graph refer to the value of $\log_{10}b_{\rm min}$ along the closest contour line.}
\label{fig1}
\end{figure}

An initial magnetic field with  strength $B_s\sim 10^{-20}$ G  could seed the galactic magnetic field assuming that in addition there is a galactic dynamo operating \cite{rees}. This leads to $r\sim 10^{-37}$ at a galactic scale of order 1 Mpc. 
The former estimate does not take into account the presence of a cosmological constant which reduces the minimal magnetic field strength to $B_s\sim 10^{-30}$  G and $r\sim 10^{-57}$ \cite{dlt}.
In order to directly  seed the magnetic field, $r$ has to be of the order of $10^{-8}$.
In the following  $r$ (cf. equation (\ref{e6})) is calculated at the galactic scale $\omega_{\rm G}=10^{-14} $Hz, corresponding to a length scale of 1 Mpc. Furthermore the physical frequency  
corresponding to the maximally amplified wave number, $k_1$, today, is given by $\omega_1(\eta_0)\simeq 6\times 10^{11}\left(\frac{H_1}{M_{\rm P}}\right)^{\frac{1}{2}}$ Hz \cite{mag-ex}.
The expression for $r$ (cf. equation (\ref{e6})) depends on the parameters $b$, $c$ and $d$ in addition to the parameters characterizing inflation $\beta$ and the Hubble parameter at the beginning of the radiation dominated stage
$\frac{H_1}{M_{\rm P}}$.  The corresponding  reheat temperature $T_1$ is given by 
$\frac{T_1}{M_{\rm P}}\sim\sqrt{\frac{H_1}{M_{\rm P}}}$.
An upper limit on $r$ can be derived by using the constraint  
\begin{eqnarray}
\mu_1\left(\frac{m_e}{H_1}\right)^{-2}>1,
\label{mu}
\end{eqnarray}
consistent with the approximation used to derive equation (\ref{e4}).
This  yields to, for $\beta\neq -2,-\frac{3}{2}, -1$, at $\omega_G=10^{-14}$ Hz
\begin{eqnarray}
r_{\rm max}(\omega_G)=10^{-79+52\nu}\left[\Gamma(\nu)\right]^2\left(\frac{1}{2}-\nu\right)^2\left(\frac{\xi_2}{4}\right)^{-\nu}\left(\frac{H_1}{M_{\rm P}}\right)^{\nu+\frac{1}{2}}.
\end{eqnarray}
For simplicity assuming that the parameters determining the curvature terms are all of the same order, so that $b\sim c\sim d$, the maximal possible ratio $r_{\rm max}$ is independent of these parameters, since $\xi_2$ in this case is given by $\xi_2=\frac{10\beta-7}{7\beta-10}$.
In figure \ref{fig1} {(\it left)} the contour lines for $\log_{10} r_{\rm max}$ are shown in this case 
as a function of $\beta$ and $\log_{10}\left(\frac{H_1}{M_{\rm P}}\right)$.
The constraint on $\mu_1$ (cf. equation (\ref{mu})) implies  in the case $b\sim c\sim d$ 
a constraint on the parameter $b$,
\begin{eqnarray}
b>b_{\rm min}\equiv \frac{10^{-45}}{\beta(7\beta-10)}\left(\frac{H_1}{M_{\rm P}}\right)^{-2}.
\end{eqnarray} 
$\log b_{\rm min}$ is shown in figure \ref{fig1} {(\it right)}. 

\begin{figure}[ht]
\centerline{
\epsfxsize=2.4in\epsfbox{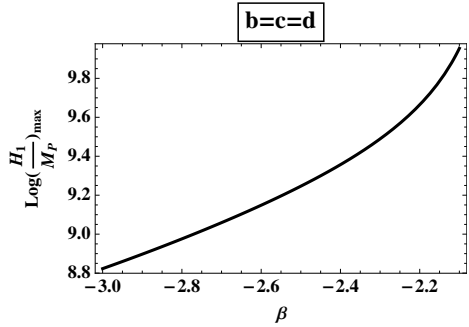}
}
\caption{The maximal value  $\log_{10}\left(\frac{H_1}{M_{\rm P}}\right)_{\rm max}$ (cf. equation (\ref{gw}))
allowed by the bound from gravitational wave production is shown for the 
model with $b=c=d$ for $h_0=0.73$.}
\label{fig2}
\end{figure}

In \cite{cd} a strong bound on primordial magnetic fields created before nucleosynthesis has been derived. The bound is due to the conversion of magnetic field energy into gravitational wave energy. Translating the bound on the magnetic field strength given in \cite{cd} 
into a bound on the magnetic field to background radiation energy density ratio $r$  results in 
\begin{eqnarray}
r_{\rm GW}\simeq 2\times 10^{-61+52\nu}{\cal N}_{\rm GW}h_0^2\left(\frac{H_1}{M_{\rm P}}\right)^{\nu-\frac{3}{2}},
\end{eqnarray}
at the galactic scale used here, $\lambda=1$ Mpc, and
\begin{eqnarray}
{\cal N}_{GW}=2^{\frac{5}{2}-\nu}\,\Gamma\left(\frac{5}{2}-\nu\right).
\end{eqnarray}
Therefore, the requirement $r_{\rm max}\leq r_{\rm GW}$ implies an upper limit on $\frac{H_1}{M_{\rm P}}$ , namely,
\begin{eqnarray}
\log_{10}\left(\frac{H_1}{M_{\rm P}}\right)_{\rm max}\equiv 9+\frac{1}{2}\log_{10}\left[\frac{2^{\frac{7}{2}-\nu}h_0^2\Gamma\left(\frac{5}{2}-\nu\right)}{\Gamma^2(\nu)\left(\frac{1}{2}-\nu\right)^2}
\left(\frac{\xi_2}{4}\right)^{\nu}\right],
\label{gw}
\end{eqnarray}
so that  the allowed range is given by $\left(\frac{H_1}{M_{\rm P}}\right)\leq\left(\frac{H_1}{M_{\rm P}}\right)_{\rm max}$. This is plotted in figure \ref{fig2} for small values of $\beta$. 
As can be appreciated from figure \ref{fig2} the maximally allowed value for $\log_{10}\left(\frac{H_1}{M_{\rm P}}\right)$ is always much bigger than 1 for $\beta<-2$ which is the relevant range here. This can always be satisfied since it is assumed that $H_1<M_{\rm P}$.

\section{Conclusions}
\setcounter{equation}{0}
Primordial magnetic field generation has been investigated in a model where  the electromagnetic field is coupled to various curvature terms which is motivated by the form of the one loop effective action of vacuum polarization in QED in a gravitational background \cite{dh,tw}. Here the resulting magnetic field spectrum has been calculated explicitly  by matching a stage of power law inflation directly to the standard radiation dominated phase, solving the mode equation and finding the appropriate Bogoliubov coefficient. The ratio of magnetic field energy density over background radiation energy density $r$ calculated at a galactic scale of 1 Mpc has been employed to compare the model with the minimal required values necessary to seed the galactic dynamic field either directly or relying on the mechanism of a galactic dynamo. For simplicity it was assumed that the coefficients of all additional curvature terms are of the same order. 
A constraint on the parameters, which is part of the approximation used to derive the mode equation on superhorizon scale, leads to a maximum value of the magnetic field strength or equivalently the ratio $r$ which is a function of $\beta$ and $\frac{H_1}{M_{\rm P}}$.
Here $\beta$ is the exponent characterizing inflation and $H_1$ is the value of the Hubble parameter at the beginning of the radiation dominated era.
As can be seen from figure \ref{fig1} there is a region in parameter space where the resulting  magnetic field strength is  larger than $B_s>10^{-20}$ G corresponding to $r>10^{-37}$, namely, $\beta<-2.4$, $\frac{H_1}{M_{\rm P}}>10^{-18}$ bounded by the corresponding contour line.
 De Sitter inflation corresponds to $\beta=-1$. In this case there is no significant magnetic field generation since the mode functions during inflation as well as during the
 radiation dominated stage are plane waves.

Finally, it has been checked that the resulting maximum magnetic field strength satisfies the bound due to gravitational wave production \cite{cd}. Imposing this bound leads to a maximal value of $\frac{H_1}{M_{\rm P}}$ which, however, is always much larger than the maximal value considered here, as can be seen from figure \ref{fig2}.
In summary, there is a range of parameters $\beta$ and $\frac{H_1}{M_{\rm P}}$ for which 
magnetic fields are generated that are strong enough to explain the galactic magnetic field.

\section{Acknowledgements}
Financial  support by Spanish Science Ministry grants
FPA2005-04823, FIS2006-05319 and CSD2007-00042 is gratefully acknowledged.

\end{document}